\documentclass[aps,prd,twocolumn,superscriptaddress,groupedaddress,floatfix]{revtex4}

\usepackage{graphicx}  % needed for figures
\usepackage{dcolumn}   % needed for some tables
\usepackage{amsmath,bm}
\usepackage{amssymb}
\usepackage{mathrsfs}
\usepackage{xspace}

\newcolumntype{K}[1]{>{\centering\arraybackslash}p{#1}}
\newcommand{\ttbar}{\ensuremath{t\bar{t}}\xspace}
\newcommand{\ppbar}{\ensuremath{p\bar{p}}\xspace}
\newcommand{\met}{\ensuremath{{}/\!\!\!{p}_{\rm T}}\xspace}
\newcommand {\pt}       {\ensuremath{p_T}}
\newcommand {\pythia}   {\textsc{pythia}\xspace}

\newcommand {\geant}    {\textsc{geant3}\xspace}
\newcommand {\alpgen}   {\textsc{alpgen}\xspace}
\newcommand {\mcatnlo}  {\textsc{mc@nlo}\xspace}
\newcommand {\herwig}   {\textsc{herwig}\xspace}
\newcommand{\comphep}   {\textsc{comphep}\xspace}
\newcommand{\madgraph}  {\textsc{madgraph}\xspace}

\newcommand {\ie}       {\textrm{i.e.}}

\newcommand {\etal}     {\textit{et al.}}
\newcommand{\ktmin}	{\ensuremath{k^{min}_T}}

\newcommand{\wplus}     {\ensuremath{W+}jets\xspace}

\newcommand{\muplus}    {\ensuremath{\mu +}jets\xspace}
\newcommand{\mm}{\mathrm}
\begin{document}

\hspace{5.2in} \mbox{FERMILAB-PUB-16-284-E}

\title{Measurement of top quark polarization in \ttbar lepton+jets final states}
\affiliation{LAFEX, Centro Brasileiro de Pesquisas F\'{i}sicas, Rio de Janeiro, Rio de Janeiro 22290, Brazil}
\affiliation{Universidade do Estado do Rio de Janeiro, Rio de Janeiro, Rio de Janeiro 20550, Brazil}
\affiliation{Universidade Federal do ABC, Santo Andr\'e, S\~{a}o Paulo 09210, Brazil}
\affiliation{University of Science and Technology of China, Hefei 230026, People's Republic of China}
\affiliation{Universidad de los Andes, Bogot\'a, 111711, Colombia}
\affiliation{Charles University, Faculty of Mathematics and Physics, Center for Particle Physics, 116 36 Prague 1, Czech Republic}
\affiliation{Czech Technical University in Prague, 116 36 Prague 6, Czech Republic}
\affiliation{Institute of Physics, Academy of Sciences of the Czech Republic, 182 21 Prague, Czech Republic}
\affiliation{Universidad San Francisco de Quito, Quito 170157, Ecuador}
\affiliation{LPC, Universit\'e Blaise Pascal, CNRS/IN2P3, Clermont, F-63178 Aubi\`ere Cedex, France}
\affiliation{LPSC, Universit\'e Joseph Fourier Grenoble 1, CNRS/IN2P3, Institut National Polytechnique de Grenoble, F-38026 Grenoble Cedex, France}
\affiliation{CPPM, Aix-Marseille Universit\'e, CNRS/IN2P3, F-13288 Marseille Cedex 09, France}
\affiliation{LAL, Univ. Paris-Sud, CNRS/IN2P3, Universit\'e Paris-Saclay, F-91898 Orsay Cedex, France}
\affiliation{LPNHE, Universit\'es Paris VI and VII, CNRS/IN2P3, F-75005 Paris, France}
\affiliation{CEA Saclay, Irfu, SPP, F-91191 Gif-Sur-Yvette Cedex, France}
\affiliation{IPHC, Universit\'e de Strasbourg, CNRS/IN2P3, F-67037 Strasbourg, France}
\affiliation{IPNL, Universit\'e Lyon 1, CNRS/IN2P3, F-69622 Villeurbanne Cedex, France and Universit\'e de Lyon, F-69361 Lyon CEDEX 07, France}
\affiliation{III. Physikalisches Institut A, RWTH Aachen University, 52056 Aachen, Germany}
\affiliation{Physikalisches Institut, Universit\"at Freiburg, 79085 Freiburg, Germany}
\affiliation{II. Physikalisches Institut, Georg-August-Universit\"at G\"ottingen, 37073 G\"ottingen, Germany}
\affiliation{Institut f\"ur Physik, Universit\"at Mainz, 55099 Mainz, Germany}
\affiliation{Ludwig-Maximilians-Universit\"at M\"unchen, 80539 M\"unchen, Germany}
\affiliation{Panjab University, Chandigarh 160014, India}
\affiliation{Delhi University, Delhi-110 007, India}
\affiliation{Tata Institute of Fundamental Research, Mumbai-400 005, India}
\affiliation{University College Dublin, Dublin 4, Ireland}
\affiliation{Korea Detector Laboratory, Korea University, Seoul, 02841, Korea}
\affiliation{CINVESTAV, Mexico City 07360, Mexico}
\affiliation{Nikhef, Science Park, 1098 XG Amsterdam, Netherlands}
\affiliation{Radboud University Nijmegen, 6525 AJ Nijmegen, Netherlands}
\affiliation{Joint Institute for Nuclear Research, Dubna 141980, Russia}
\affiliation{Institute for Theoretical and Experimental Physics, Moscow 117259, Russia}
\affiliation{Moscow State University, Moscow 119991, Russia}
\affiliation{Institute for High Energy Physics, Protvino, Moscow region 142281, Russia}
\affiliation{Petersburg Nuclear Physics Institute, St. Petersburg 188300, Russia}
\affiliation{Instituci\'{o} Catalana de Recerca i Estudis Avan\c{c}ats (ICREA) and Institut de F\'{i}sica d'Altes Energies (IFAE), 08193 Bellaterra (Barcelona), Spain}
\affiliation{Uppsala University, 751 05 Uppsala, Sweden}
\affiliation{Taras Shevchenko National University of Kyiv, Kiev, 01601, Ukraine}
\affiliation{Lancaster University, Lancaster LA1 4YB, United Kingdom}
\affiliation{Imperial College London, London SW7 2AZ, United Kingdom}
\affiliation{The University of Manchester, Manchester M13 9PL, United Kingdom}
\affiliation{University of Arizona, Tucson, Arizona 85721, USA}
\affiliation{University of California Riverside, Riverside, California 92521, USA}
\affiliation{Florida State University, Tallahassee, Florida 32306, USA}
\affiliation{Fermi National Accelerator Laboratory, Batavia, Illinois 60510, USA}
\affiliation{University of Illinois at Chicago, Chicago, Illinois 60607, USA}
\affiliation{Northern Illinois University, DeKalb, Illinois 60115, USA}
\affiliation{Northwestern University, Evanston, Illinois 60208, USA}
\affiliation{Indiana University, Bloomington, Indiana 47405, USA}
\affiliation{Purdue University Calumet, Hammond, Indiana 46323, USA}
\affiliation{University of Notre Dame, Notre Dame, Indiana 46556, USA}
\affiliation{Iowa State University, Ames, Iowa 50011, USA}
\affiliation{University of Kansas, Lawrence, Kansas 66045, USA}
\affiliation{Louisiana Tech University, Ruston, Louisiana 71272, USA}
\affiliation{Northeastern University, Boston, Massachusetts 02115, USA}
\affiliation{University of Michigan, Ann Arbor, Michigan 48109, USA}
\affiliation{Michigan State University, East Lansing, Michigan 48824, USA}
\affiliation{University of Mississippi, University, Mississippi 38677, USA}
\affiliation{University of Nebraska, Lincoln, Nebraska 68588, USA}
\affiliation{Rutgers University, Piscataway, New Jersey 08855, USA}
\affiliation{Princeton University, Princeton, New Jersey 08544, USA}
\affiliation{State University of New York, Buffalo, New York 14260, USA}
\affiliation{University of Rochester, Rochester, New York 14627, USA}
\affiliation{State University of New York, Stony Brook, New York 11794, USA}
\affiliation{Brookhaven National Laboratory, Upton, New York 11973, USA}
\affiliation{Langston University, Langston, Oklahoma 73050, USA}
\affiliation{University of Oklahoma, Norman, Oklahoma 73019, USA}
\affiliation{Oklahoma State University, Stillwater, Oklahoma 74078, USA}
\affiliation{Oregon State University, Corvallis, Oregon 97331, USA}
\affiliation{Brown University, Providence, Rhode Island 02912, USA}
\affiliation{University of Texas, Arlington, Texas 76019, USA}
\affiliation{Southern Methodist University, Dallas, Texas 75275, USA}
\affiliation{Rice University, Houston, Texas 77005, USA}
\affiliation{University of Virginia, Charlottesville, Virginia 22904, USA}
\affiliation{University of Washington, Seattle, Washington 98195, USA}
\author{V.M.~Abazov} \affiliation{Joint Institute for Nuclear Research, Dubna 141980, Russia}
\author{B.~Abbott} \affiliation{University of Oklahoma, Norman, Oklahoma 73019, USA}
\author{B.S.~Acharya} \affiliation{Tata Institute of Fundamental Research, Mumbai-400 005, India}
\author{M.~Adams} \affiliation{University of Illinois at Chicago, Chicago, Illinois 60607, USA}
\author{T.~Adams} \affiliation{Florida State University, Tallahassee, Florida 32306, USA}
\author{J.P.~Agnew} \affiliation{The University of Manchester, Manchester M13 9PL, United Kingdom}
\author{G.D.~Alexeev} \affiliation{Joint Institute for Nuclear Research, Dubna 141980, Russia}
\author{G.~Alkhazov} \affiliation{Petersburg Nuclear Physics Institute, St. Petersburg 188300, Russia}
\author{A.~Alton$^{a}$} \affiliation{University of Michigan, Ann Arbor, Michigan 48109, USA}
\author{A.~Askew} \affiliation{Florida State University, Tallahassee, Florida 32306, USA}
% ...
%
% visitor_addresses.tex                       1 November 2016
%  available symbols are:
%  $\ast, \dag, \ddag, \S, \P, $\|$, $\ast\ast$, \dag\dag, \ddag\ddag ,\#
%
\collaboration{The D0 Collaboration\footnote{with visitors from
%{alton}
$^{a}$Augustana College, Sioux Falls, SD 57197, USA,
%{burdin}
$^{b}$The University of Liverpool, Liverpool L69 3BX, UK,
%{grohsjean,deterre}
$^{c}$Deutshes Elektronen-Synchrotron (DESY), Notkestrasse 85, Germany,
%{de la cruz-burelo}
$^{d}$CONACyT, M-03940 Mexico City, Mexico,
%{partridge}
$^{e}$SLAC, Menlo Park, CA 94025, USA,
%{hesketh}
$^{f}$University College London, London WC1E 6BT, UK,
%{luna-garcia}
$^{g}$Centro de Investigacion en Computacion - IPN, CP 07738 Mexico City, Mexico,
%{santos}
$^{h}$Universidade Estadual Paulista, S\~ao Paulo, SP 01140, Brazil,
%{meyer}
$^{i}$Karlsruher Institut f\"ur Technologie (KIT) - Steinbuch Centre for Computing (SCC),
D-76128 Karlsruhe, Germany,
%{patwa}
$^{j}$Office of Science, U.S. Department of Energy, Washington, D.C. 20585, USA,
%{cooke}
$^{k}$American Association for the Advancement of Science, Washington, D.C. 20005, USA,
%{borysova}
$^{l}$Kiev Institute for Nuclear Research (KINR), Kyiv 03680, Ukraine,
%{jabeen}
$^{m}$University of Maryland, College Park, MD 20742, USA,
%{williams}
$^{n}$European Orgnaization for Nuclear Research (CERN), CH-1211 Geneva, Switzerland,
%{Jung}
$^{o}$Purdue University, West Lafayette, IN 47907, USA,
and
%{Zivkovic}
$^{p}$Institute of Physics, Belgrade, Belgrade, Serbia.
%{montgomery}
%$^{?}$Thomas Jefferson National Accelerator Facility, Newport News, VA 23606, USA,
%{falkowski}
%$^{?}$Laboratoire de Physique Theorique, F-91405 Orsay CEDEX, FR,
%{hooper,kozminski}
%$^{?}$}Visitor from Lewis University, Romeoville, IL 60446, USA.
%{weber}
%$^{?}$Universit{\"a}t Bern, CH-3012 Bern, Switzerland.
%{deceased}
%{peng, lipaev, cihangir}
$^{\ddag}$Deceased.
}} \noaffiliation
\vskip 0.25cm

\date{July 27, 2016}

\begin{abstract}
We present a measurement of top quark polarization in \ttbar pair production in \ppbar collisions at $\sqrt{s}=1.96$\,TeV using data corresponding to 9.7\,fb$^{-1}$ of integrated luminosity recorded with the D0 detector at the Fermilab Tevatron Collider. We consider final states containing a lepton and at least three jets. The polarization is measured through the distribution of lepton angles along three axes: the beam axis, the helicity axis, and the transverse axis normal to the \ttbar production plane. This is the first measurement of top quark polarization at the Tevatron using lepton+jet final states and the first measurement of the transverse polarization in $t\overline t$ production. The observed distributions are consistent with standard model predictions of nearly no polarization. 
\end{abstract}

\maketitle

\section{Introduction}

The standard model (SM) predicts that top quarks produced at the Tevatron collider are almost unpolarized,
while models beyond the standard model (BSM) predict enhanced polarizations~\cite{Fajfer:2012si}.
The top quark polarization $P_{\hat{n}}$ can be measured in the top quark rest frame through the angular distributions of the top quark decay
products relative to some chosen axis $\hat{n}$~\cite{Bernreuther},
\begin{equation}
\frac{1}{\Gamma}\frac{d\Gamma}{d\cos{\theta_{i,\hat{n}}}}=\frac{1}{2}(1+P_{\hat{n}}\kappa_{i}\cos{\theta_{i,\hat{n}}}),
\label{eqn:0}
\end{equation}
where $i$ is the decay product (lepton, quark, or neutrino), $\kappa_{i}$ is its spin-analyzing power ($\approx 1$ for charged leptons, 0.97 for $d$-type quarks, $-0.4$ for $b$-quarks, and $-0.3$ for neutrinos and $u$-type quarks ~\cite{Brandenburg}), and $\theta_{i,\hat{n}}$ is
the angle between the direction of the decay product $i$ and the quantization axis $\hat{n}$. 
The mean polarizations of the top and antitop quarks are expected to be identical because of $CP$ conservation.
The $P_{\hat{n}}$ can be obtained from the asymmetry of the $\cos{\theta}$ distribution 
\begin{equation}
A_{P, \hat{n}}=\frac{N(\cos{\theta_{i,\hat{n}}}>0)-N(\cos{\theta_{i,\hat{n}}}<0)}{N(\cos{\theta_{i,\hat{n}}}>0)+N(\cos{\theta_{i,\hat{n}}}<0)},
\label{eqn:m1}
\end{equation}
where $N(x)$ is the number of events passing the requirement $x$ and the polarization is then  $P_{\hat{n}} = 2 A_{P, \hat{n}}$.
The quantization axes are defined in the \ttbar rest frame, while the decay product directions are defined after successively boosting the particles to the \ttbar rest frame and then to the parent top quark rest frame. 
We measure the polarization along three quantization axes: \textrm{(i)} the {\bf beam axis $\hat{n}_{p}$}, given by the direction of the proton beam~\cite{Bernreuther}; \textrm{(ii)} the {\bf helicity axis $\hat{n}_{h}$}, given by the direction of the parent top or antitop quark; and the \textrm{(iii)} {\bf transverse axis $\hat{n}_T$}, given as perpendicular to the production plane defined by the proton and parent top quark directions, \ie ,
$\hat{n}_{p} \times \hat{n}_{t}$ (or by $\hat{n}_{p} \times - \hat{n}_{\overline{t}}$ for the antitop quark)~\cite{Bernreuther:1995cx, Baumgart:2013yra}.

The D0 Collaboration published a short study of the top quark polarization along the helicity axis in \ppbar collisions as part of the measurement of angular asymmetries of leptons~\cite{tevpol}, but no measured value was presented. Recently, the D0 Collaboration measured the top quark polarization along the beam axis in \ttbar final states with two leptons \cite{Boris}, finding it to be consistent with the SM. The ATLAS and CMS collaborations measured the top quark polarization along the helicity axis in $pp$ collisions, and the results are consistent with no polarization \cite{atlaspol, cmspol}. The polarization at the Tevatron and LHC are expected to be different because of the difference in the initial states, which motivates the measurement of the polarizations in Tevatron data~\cite{top2014,Choudhury:2010cd}. For beam and transverse axes, the top quark polarizations in \ppbar collisions are expected to be larger than those for $pp$~\cite{Bernreuther,Bernreuther:1995cx}, therefore offering greater sensitivity to BSM models with nonzero polarization.

The longitudinal polarizations along the beam and helicity axes at the Tevatron collider are predicted by the SM to be $(-0.19 \pm 0.05) \%$ and $(-0.39 \pm 0.04) \%$~\cite{weakcorr}, respectively, while the transverse polarization is estimated to be $\approx 1.1 \%$~\cite{Baumgart:2013yra}.
Observation of a significant departure from the expected value would be evidence for BSM contributions to the top quark polarization~\cite{Fajfer:2012si}.

% We present a measurement of top quark polarization in $\ell$+jets final states of \ttbar production.
We present a measurement of top quark polarization in $\ell$+jets final states of \ttbar production using data collected with the D0 detector~\cite{run2det}, corresponding to an integrated luminosity of 9.7\,fb$^{-1}$ of \ppbar collisions at $\sqrt{s}=1.96$\,TeV. The lepton is most sensitive to the polarization and is easily identified. We therefore examine the angular distribution of leptons. 
After selecting the events in the $\ell$+jets final state, we perform a kinematic fit to reconstruct the lepton angles relative to the various axes.
The resulting distributions are fitted with mixtures of signal templates with $+1$ and $-1$ polarizations to extract the observed values.
The down-type quark has an analyzing power close to unity, but its identification is difficult. It is therefore not used in the measurement. 
However, to gain statistical precision we use reweighted Monte Carlo (MC) down-type quark distributions in forming signal event templates.

\section{Event selection}

Each top quark of the \ttbar pair decays into a $b$ quark and a $W$ boson with nearly $100\%$ probability, leading to a $W^+W^-b\bar b$ final state. In $\ell$+jets events, one of the $W$ bosons decays leptonically and the other into quarks that evolve into jets. 
The trigger selects $\ell$+jets events with at least one lepton, electron~($e$) or a muon~($\mu$). The efficiency of the trigger is $95 \%$ or $80 \%$ for \ttbar events containing reconstructed $e$ or $\mu$ candidates, respectively.
This analysis requires the presence of one isolated $e$~\cite{bib:emid} or $\mu$~\cite{bib:muid} with transverse momentum $\pt>20$\,GeV and physics pseudorapidity~\cite{foot} $|\eta|<1.1$ or $|\eta|<2$, respectively. In addition, leptons are required to originate from within 1\,cm of the primary \ppbar interaction vertex (PV) in the coordinate along the beam axis.
Accepted events must have a reconstructed PV within $60$\,cm of the center of the detector along the beam axis. Furthermore, we require an imbalance in transverse momentum $\met>20$\,GeV, expected from the undetected neutrino.
Jets are reconstructed using an iterative cone algorithm~\cite{bib:cone} with a cone parameter of $R=0.5$.
Jet energies are corrected to the particle level using calibrations from studies of exclusive $\gamma+$jet, $Z+$jet, and dijet events~\cite{bib:jes}. These calibrations account for differences in the detector response to jets originating from gluons, $b$~quarks, and $u,d,s,$ or $c$~quarks. 
We require at least three jets with $\pt>20$\,GeV within $|\eta|<2.5$, and $\pt>40$\,GeV for the jet of highest \pt. 
At least one jet per event is required to be identified as originating from a $b$ quark ($b$ tagged) through the use of a multivariate algorithm~\cite{bib:bid}. 
In \muplus events, upper limits are required on the transverse mass of the reconstructed $W$ boson~\cite{Smith:1983aa} of $M_T^W < 250$\,GeV and \met$<250$\,GeV to remove events with misreconstructed muon $p_T$. Additional selections are applied to reduce backgrounds in muon events, and to suppress contributions from multijet production. A detailed description of these requirements can be found in Ref.~\cite{bib:diffxsec}. In addition, we require the curvature of the track associated with the lepton to be well measured to reduce lepton charge misidentification.

\section{Signal and background samples}

We simulate \ttbar events at the next-to-leading-order (NLO) in perturbative QCD with the \mcatnlo event generator version 3.4 \cite{mcatnlo} and at the leading-order (LO) with \alpgen event generator version 2.11 \cite{alpgen}. Parton showering, hadronization, and modeling of the underlying event are performed with \herwig \cite{herwig} for \mcatnlo events and with \pythia 6.4 \cite{pythia} for \alpgen events. The detector response is simulated using \geant \cite{geant}. To model the effects of multiple \ppbar interactions, the MC events are overlaid with events from random \ppbar collisions with the same luminosity distribution as the data. The main background to the \ttbar signal is \wplus events, where the $W$ boson is produced via the electroweak interaction together with additional partons from QCD radiation. The \wplus final state can be split into four subsamples according to parton flavor, $Wb\bar{b}+\mm{jets}$, $Wc\bar{c}+\mm{jets}$, $Wc+\mm{jets}$, and $W+$light jets, where light refers to gluons, $u$, $d$, or $s$ quarks. The \wplus background is modeled with \alpgen and \pythia \cite{alpgen,pythia}, as is the background from $Z+$jets events. Other background processes include $WW$, $WZ$, and $ZZ$ diboson productions simulated using \pythia, and single top quark electroweak production simulated using \comphep \cite{comphep}.
The multijet background, where a jet is misidentified as an isolated lepton, is estimated from the data using the matrix method \cite{matrixMethod,bib:diffxsec}. 
We use six different BSM models~\cite{Axi} to study modified \ttbar production: one $Z'$ boson model and five axigluon models with different axigluon masses and couplings (m200R, m200L, m200A, m2000R, and m2000A, where L, R, and A refer to left-handed, right-handed, and axial couplings, and numbers are the particle masses in GeV). Some additional axigluon models such as m2000L are not simulated as they are excluded by other measurements of top quark properties.
The BSM events are generated with LO \madgraph 5 \cite{madgraph} interfaced to \pythia for parton evolution.

\section{Analysis method}

A constrained kinematic $\chi ^{2}$ fit is used to associate the observed leptons and jets with the individual top quarks using a likelihood term for each jet-to-quark assignment, as described in Ref.~\cite{Abazov:2014cca}. We assume the four jets with largest $p_T$ to originate from \ttbar decay in events with more than four jets. The algorithm includes a technique that reconstructs events with a lepton and only three jets~\cite{rocfit}.
The addition of the three-jet sample almost doubles the signal sample as shown in Table~\ref{Tab:Sample}.
In our analysis, all possible assignments of jets to final state quarks are considered and weighted by the $\chi ^{2}$ probability of each kinematic fit and by the $b$ tagging probability.

To determine the sample composition, we construct a kinematic discriminant based on the approximate likelihood ratio of expectations for \ttbar and \wplus events~\cite{Abazov:2007kg}. The input variables are chosen to achieve good separation between \ttbar and \wplus events, and required to be well modeled and not strongly correlated with one another or with the lepton polar angles used in the measurement.
Sets of input variables are selected independently for the $\ell +3$ jet and the $\ell + \geq 4$ jet events, each in three subchannels according to the number of $b$ tagged jets: 0, 1, $\geq 2$. The channels without $b$ tagged jets are used to determine the sample composition and background calibration, not to measure the polarization.

The input variables used for the $\ell$+3 jet kinematic discriminant are \ktmin = min$(p_{T,a},p_{T,b}) \cdot \Delta R_{ab}$, where $\Delta R_{ab} = \sqrt{(\eta_a - \eta_b)^2 + (\phi_a - \phi_b)^2}$ is the angular distance between the two closest jets ($a$ and $b$),  min$(p_{T,a},p_{T,b})$ represents the smaller transverse momentum of the two jets, and the $\phi$ are their azimuths in radians; aplanarity $A = 3/2 \lambda _3$, where $\lambda _3$ is the smallest eigenvalue of the normalized momentum tensor; $H_{T}^{\ell}$, which is the scalar sum of the $p_T$ of the jets and lepton; $\Delta R$ between the leading jet and the next-to-leading jet; and $\Delta R$ between the lepton and the leading jet.

The input variables for the $\ell + \geq 4$ jet discriminant are \ktmin; aplanarity; $H_{T}^{\ell}$; centrality, $C = H_T / H$, where $H_T$ is the scalar sum of all jet $p_T$ values and $H$ is the scalar sum of all jet energies; the lowest $\chi ^{2}$ among the different kinematic fit solutions in each event; $(p_{T}^{b_\mathrm{had}}-p_{T}^{b_\mathrm{lep}})/(p_{T}^{b_\mathrm{had}}+p_{T}^{b_\mathrm{lep}})$, the relative $p_T$ difference between $b_\mathrm{lep}$, the $b$ jet candidate from the $t \rightarrow b \ell \nu$ decay, and $b_\mathrm{had}$, the $b$ jet candidate from the $t \rightarrow bqq'$ decay; and $m_{jj}$, the invariant mass of the two jets corresponding to the $W \rightarrow q \overline{q}'$ decay.

The sample composition is determined from a simultaneous maximum-likelihood fit to the kinematic discriminant distributions.
The \wplus background is normalized separately for the heavy-flavor contribution ($Wb\bar{b}+\mm{jets}$ and $Wc\bar{c}+\mm{jets}$) and for the light-parton contribution ($Wc+\mm{jets}$ and $W+$light jets). The sample composition after implementing the selections, and fitting the maximum likelihood to data, is broken down into individual channels by lepton flavor and number of jets, and summarized in Table~\ref{Tab:Sample}. The obtained \ttbar yield is close to the expectations.
\begin{table}[h!]
\begin{center}
\begin{ruledtabular}\begin{tabular}{lcccc}
          & \multicolumn{2}{c}{3 jets} & \multicolumn{2}{c}{$\geq 4$ jets} \\
Source    & $e$+jets & $\mu$+jets & $e$+jets & $\mu$+jets \\ \hline
$W$+jets          &  $1741 \pm 26$       & $1567 \pm 15$         &  $339 \pm 3$       &  $295 \pm 3$        \\
Multijet        &  $494 \pm 7$        & $128 \pm 3$          &  $147 \pm 4$         &  $49 \pm 2$        \\
Other Bkg       &  $446 \pm 5$       & $378 \pm 2$          &  $87 \pm 1$        &  $73 \pm 1$        \\
$t\overline{t}$ signal &  $1200 \pm 25$       & $817 \pm 20$         &  $1137 \pm 24$      &  $904 \pm 23$        \\ \hline
Sum      &  $3881 \pm 37$      & $2890 \pm 25$        &  $1710 \pm 25$      &  $1321 \pm 23 $        \\ \hline
Data            &  $3872$       &  $2901$        &  $1719$       &    $1352$      \\
\end{tabular}\end{ruledtabular}
\end{center}
\caption{
 Sample composition and event yields after implementing the selection requirements and the maximum-likelihood fit to kinematic distributions in data. Only statistical uncertainties are shown.
}
\label{Tab:Sample}
\end{table}

\begin{figure}[h!]
\begin{centering}
\includegraphics[width=0.99\columnwidth]{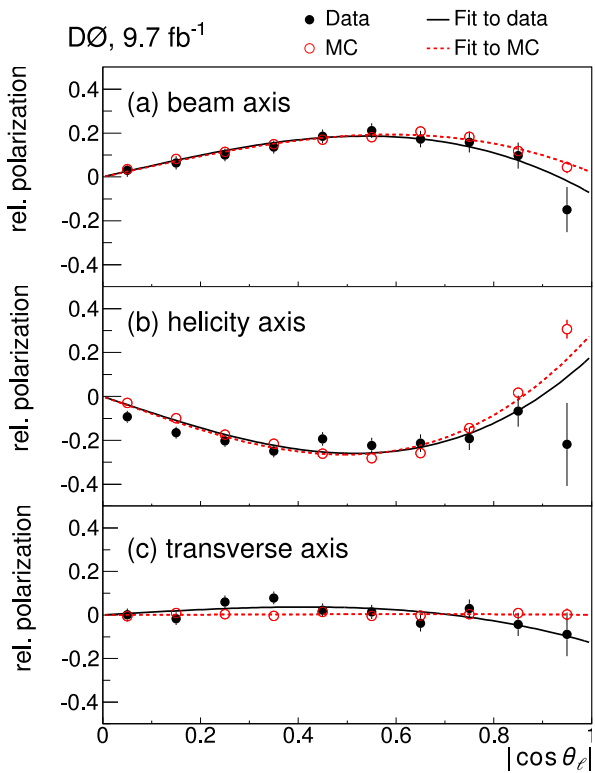}\\
\par\end{centering}
\caption{
\label{fig:wjc}
The simulated \wplus events before correction and data with \ttbar and other than \wplus background components subtracted compared in $\cos{\theta_{\ell,\hat{n}}}$ distributions in the $\ell$+3 jet and no $b$ tagged jet control sample.
% Panel (a) represents the beam axis, (b) the helicity axis, and (c) the transverse axis.
}
\end{figure}

The lepton angular distributions in $W+$jets events must be well modeled since these events form the leading background, especially in the $\ell$+3 jet sample. We therefore use a control sample of $\ell$+3 jet events without $b$ tagged jets, as such events are dominated by $W$+jets production with $>70\%$ contribution. This sample is not used for the polarization measurement.
We reweight the \wplus MC events so that the $\cos{\theta_{\ell,\hat{n}}}$ distributions agree with those for the control events in data with \ttbar and other background components subtracted. We use the relative polarization asymmetry defined as $[N_{j}(\cos{\theta_{l,\hat{n}}})-N_{-j}(\cos{\theta_{l,\hat{n}}})]/[N_{j}(\cos{\theta_{l,\hat{n}}})+N_{-j}(\cos{\theta_{l,\hat{n}}})]$, where $j$ refers to bins of $\cos{\theta_{\ell,\hat{n}}}$ values between 0 and 1 and $-j$ refers to bins between $-1$ and 0. The distributions of simulated \wplus events and subtracted data are shown in Fig.~\ref{fig:wjc}.
The correction to MC obtained from the control sample is applied to the background templates used in our signal extraction. The corrections are $0.047 \pm 0.002$ for polarization along the beam axis, $0.011 \pm 0.001$ for the transverse axis, and a negligible amount for the helicity axis. The uncertainties are propagated to the measurement as a systematic uncertainty of the background modeling.
We observe the $W+$jets events to have polarization, calculated as in Eq.~(\ref{eqn:m1}), of $+$0.18 along the beam axis, $-$0.23 along the helicity axis, and $-$0.02 along the transverse axis. Other backgrounds give polarizations of $+$0.05 (beam axis), $-$0.30 (helicity axis), and $+$0.01 (transverse axis).  

To measure the polarization, a fit is performed to the reconstructed $\cos{\theta_{\ell,\hat{n}}}$ distribution using \ttbar templates of $+1$ and $-1$ polarizations, and background templates normalized to the expected event yield. The signal templates arise from the \ttbar MC sample generated with no polarization but reweighted to follow the expected double differential distribution~\cite{Bernreuther},
\begin{multline}
\frac{1}{\Gamma} \frac{d\Gamma}{d\cos{\theta_{1}}\cos{\theta_{2}}} = \frac{1}{4}(1+\kappa_{1}P_{\hat{n},1}\cos{\theta_{1}}+ \\ +\rho\kappa_{2}P_{\hat{n},2}\cos{\theta_{2}}-\kappa_{1}\kappa_{2}C\cos{\theta_{1}}\cos{\theta_{2}}),
\label{Eq.RW}
\end{multline}
where indices 1 and 2 represent the $t$ and $\bar{t}$ quark decay products (the leptons and down quarks, or their charge conjugates), $\kappa$ is the spin-analyzing power, and $C$ is the \ttbar spin correlation coefficient for a given quantization axis.
We use the SM values $C=-0.368$ (helicity axis) and $C=0.791$ (beam axis), both calculated at NLO in QCD and in electroweak couplings in Ref.~\cite{Bernreuther}. The spin correlation factor is not known for the transverse axis, and thus we set $C=0$ and assign a systematic uncertainty by varying the choice of this factor. The $P_{\hat{n},i}$ represents the polarization state we model (here $P_{\hat{n},i} = \pm 1$) along the chosen axis $\hat{n}$. 
In the SM, assuming $CP$ invariance, $P_{\hat{n},1} =  P_{\hat{n},2}$ and gives the relative sign factor $\rho$ a value of +1 for the helicity axis and $-1$ for the beam and transverse axes \cite{Bernreuther}.

\begin{figure}[h!]
\begin{centering}
\includegraphics[width=0.95\columnwidth]{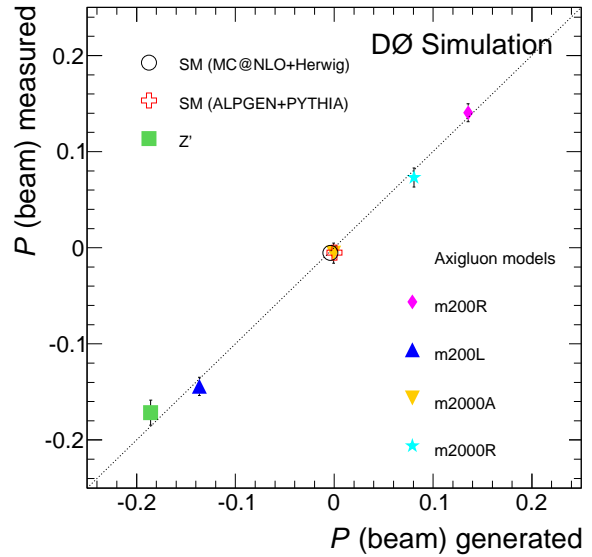}\\
\par\end{centering}
\caption{
\label{fig:closure}
Comparison of measured and generated polarizations along the beam axis for the SM and several non-SM models. The uncertainties are statistical. 
}
\end{figure}

\begin{figure*}[ht]
\begin{centering}
\includegraphics[width=0.68\columnwidth]{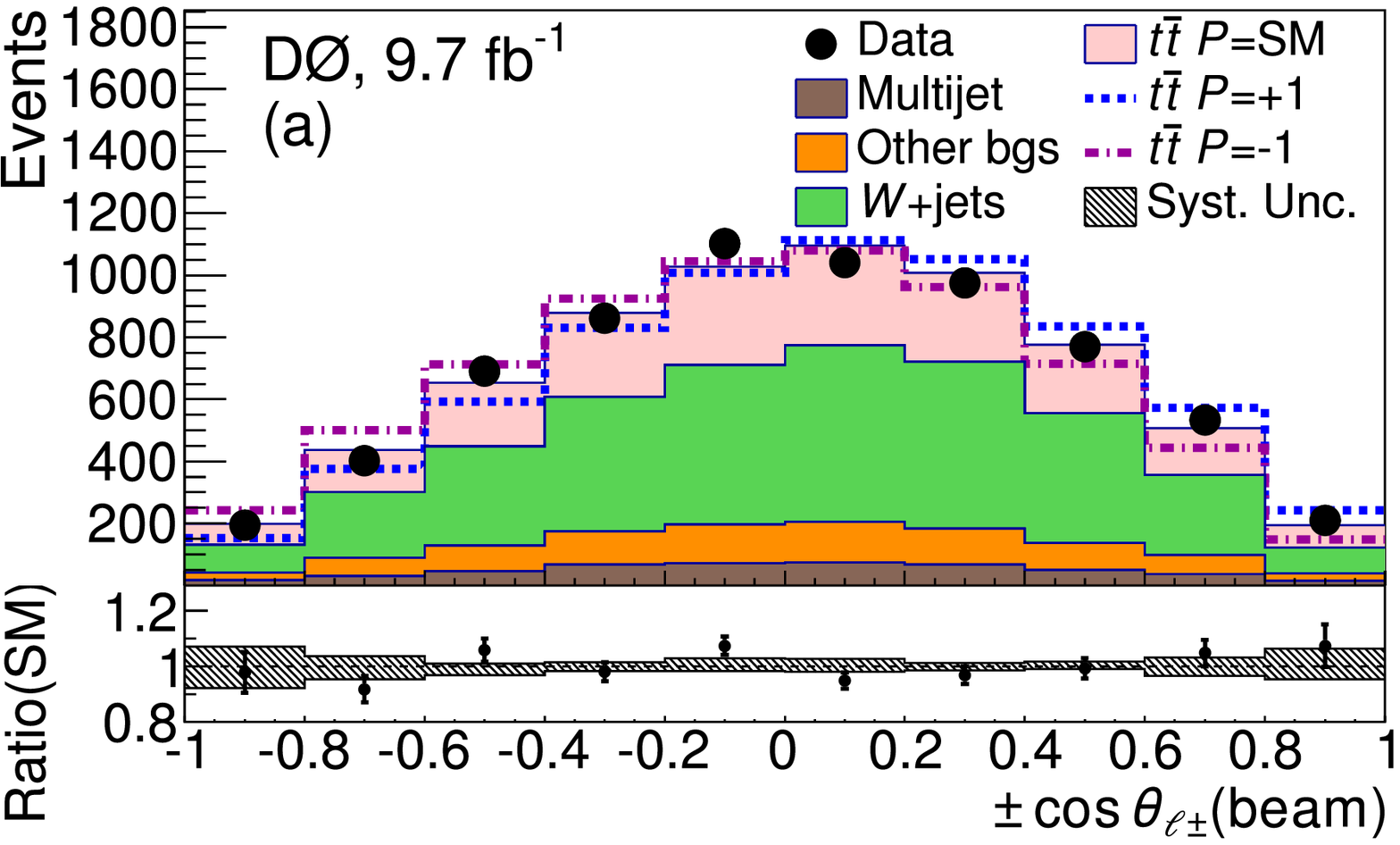}
\includegraphics[width=0.68\columnwidth]{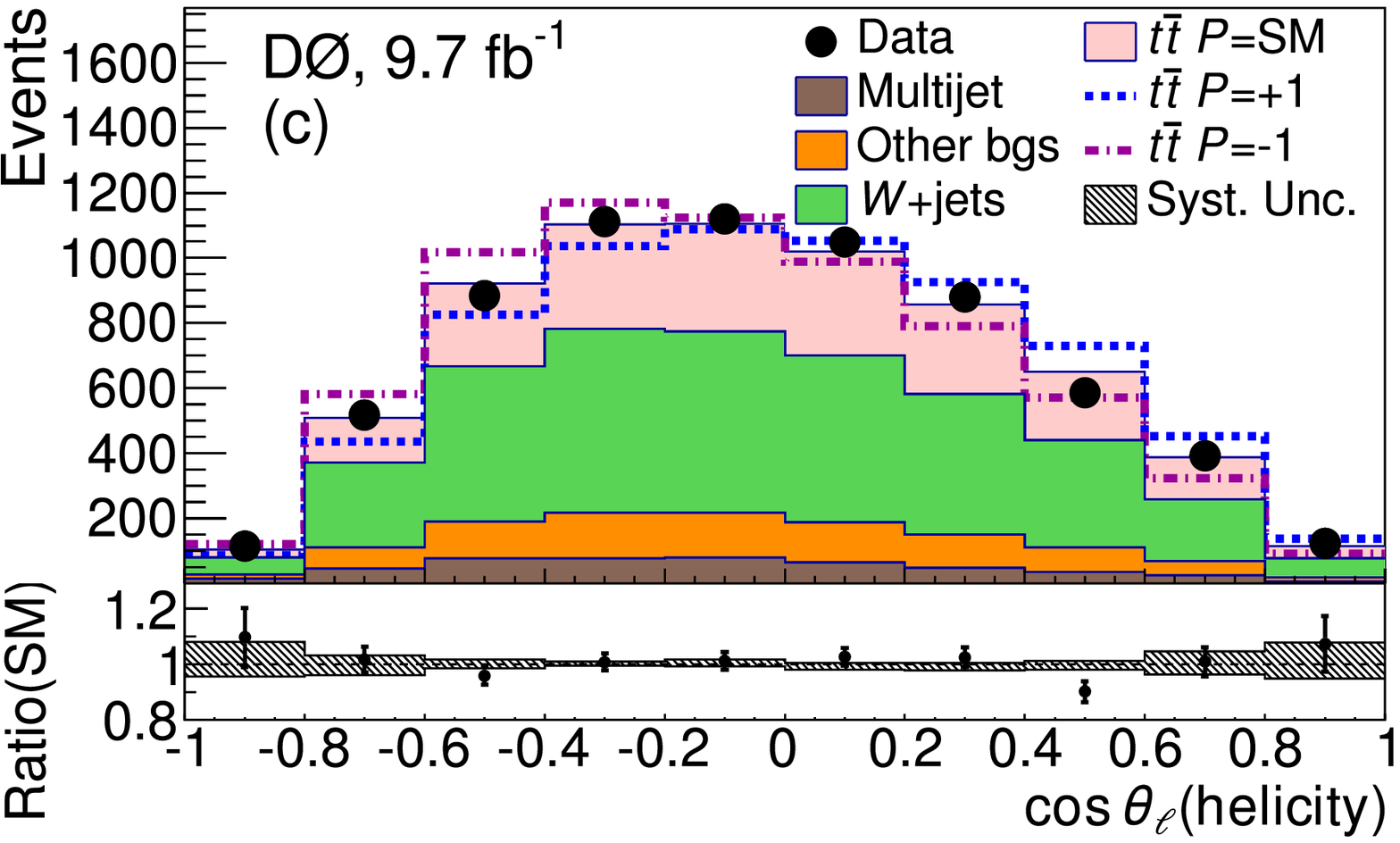}
\includegraphics[width=0.68\columnwidth]{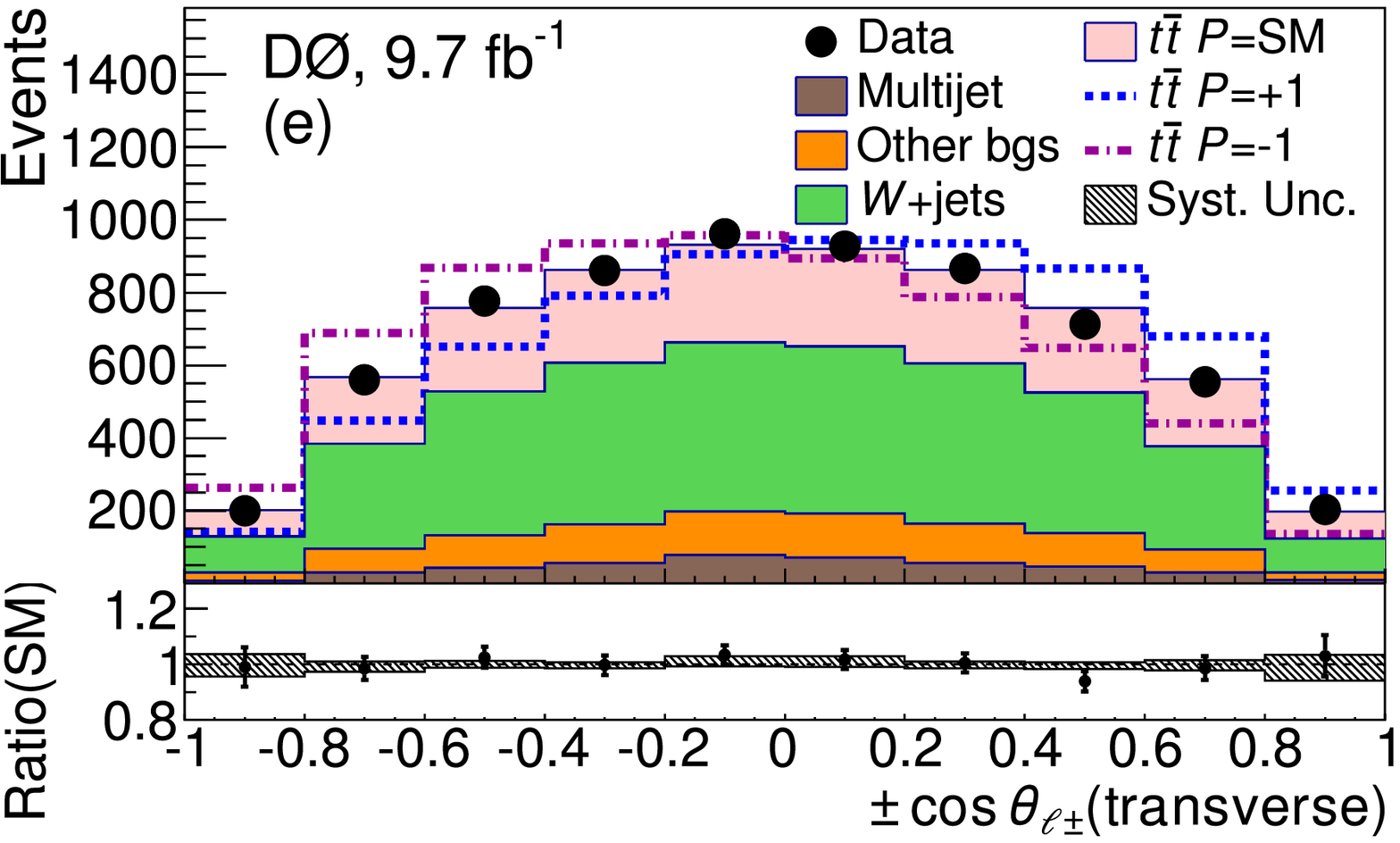}\\
\includegraphics[width=0.68\columnwidth]{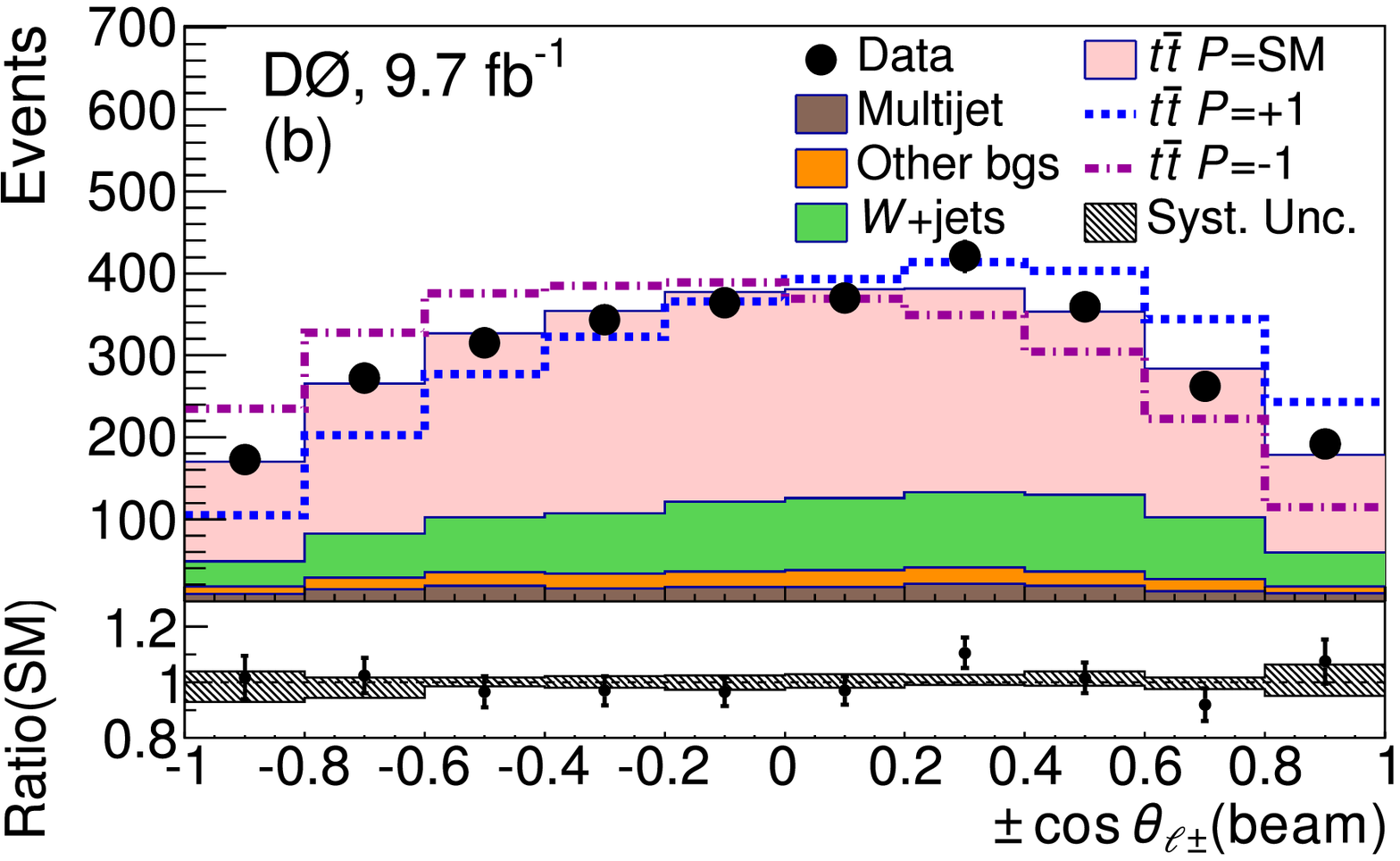} 
\includegraphics[width=0.68\columnwidth]{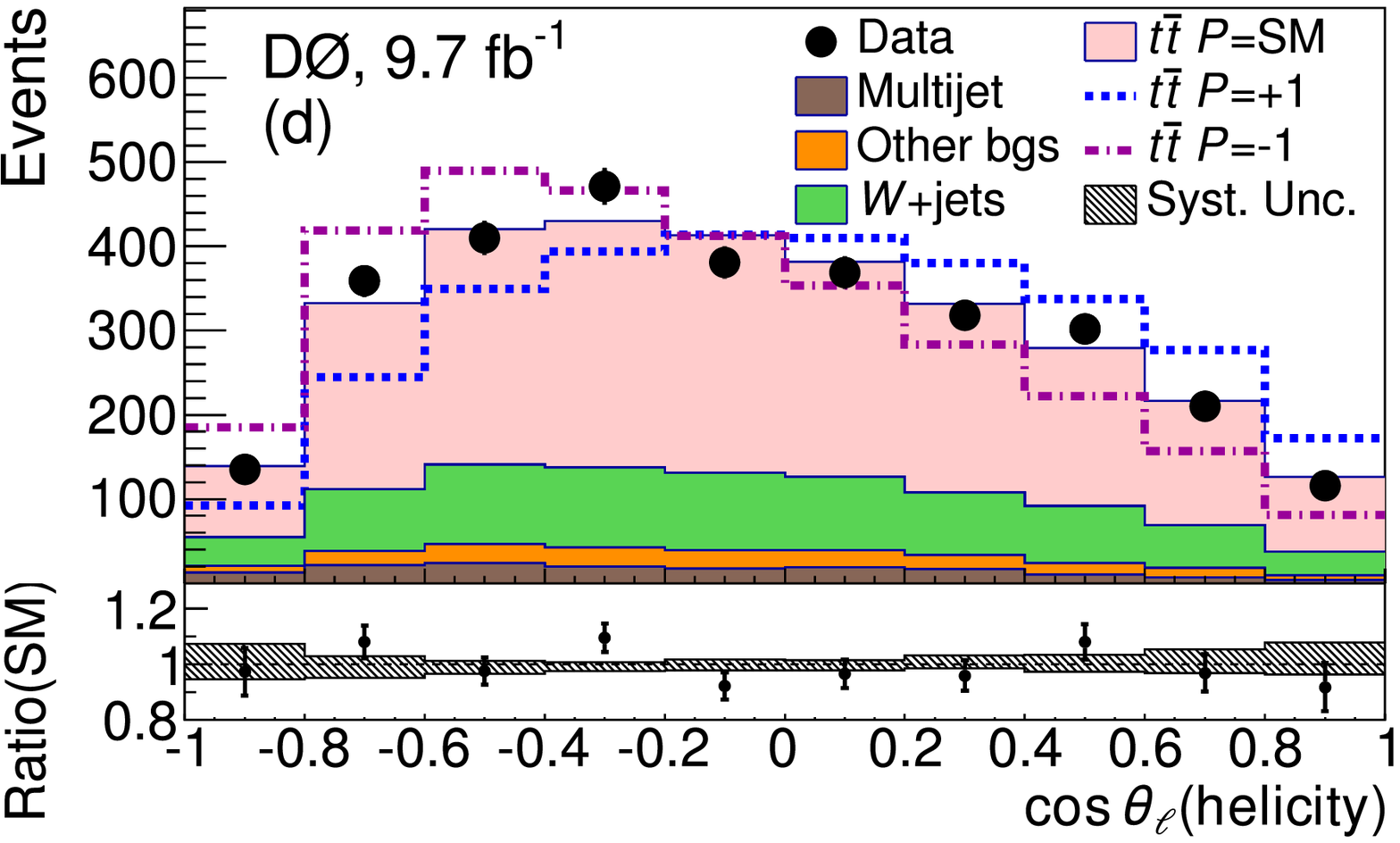} 
\includegraphics[width=0.68\columnwidth]{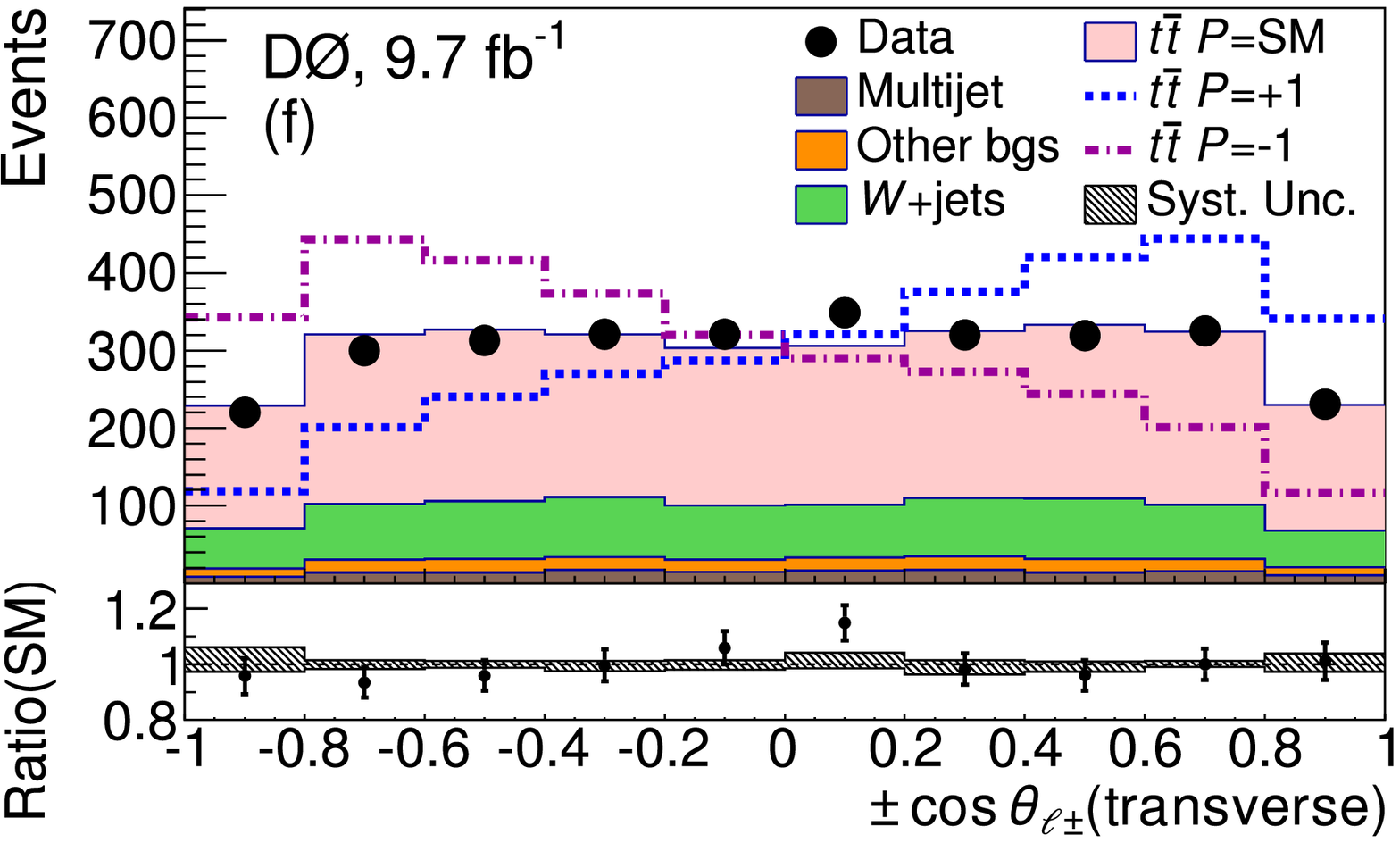}\\
\par\end{centering}
\caption{
\label{fig:templatefit}
The combined $e+$jets and $\mu+$jets $\cos\theta$ distributions for data, expected backgrounds, and signal templates for $P=-1$, SM, and +1.
Panels (a), (c), and (e) show $\ell$+3 jet events; (b), (d), and (f) show $\ell + \geq 4$ jet events; 
(a) and (b) show distributions relative to the beam axis;
(c) and (d) show distributions relative to the helicity axis;
and (e) and (f) show distributions relative to the transverse axis.
The hashed areas represent systematic uncertainties. The direction of the $\cos\theta$ axis is reversed for the $\ell^-$ events for beam and transverse spin-quantization axes plots.
}
\end{figure*}

A simultaneous fit is performed for the eight samples defined according to lepton flavor ($e$ or $\mu$), lepton charge, and number of jets (3 or $\geq 4$). The observed polarization is taken as $P=f_+ - f_-$, where $f_\pm$ are the fraction of events with $P=+1$ and $-1$ returned from the fit. The fitting procedure and methodological approach are verified using pseudoexperiments for five values of polarization, and through a check of consistency with predictions, using the BSM models with nonzero generated longitudinal polarizations.
The fitted polarizations and the model inputs are in good agreement, as shown in Fig.~\ref{fig:closure} for the polarizations along the beam axis, thus verifying our template methodology.
The distributions in the cosine of the polar angle of leptons from \ttbar decay for all three axes are shown in Fig.~\ref{fig:templatefit}.

A previous measurement of top quark polarization and the forward-backward $t$ and $\bar{t}$ asymmetry in dilepton final states~\cite{Boris} noted a correlation between these two measurements. This correlation is caused by acceptance and resolution effects in the kinematic reconstruction of the events. 
We determine the dependence of the observed polarization on the forward-backward asymmetry at the parton level, $A_{\mathrm{FB}}$, using samples in which the $t$ and $\bar t$ rapidity distributions are reweighted to accommodate the polarizations.
We then use a correction for the difference between the nominal \mcatnlo production-level $A_{\mathrm{FB}}$ of $(5.01 \pm 0.03) \%$ and the next-to-next-to-leading-order (NNLO) calculation \cite{AfbNNLO} of $(9.5 \pm 0.7) \%$. The observed correction is $-0.030$ for the polarization along the beam axis, less than $0.002$ for the polarization along the helicity axis, and is negligible for the transverse polarization. The uncertainty on the expected $A_{\mathrm{FB}}$ is propagated to the measurement as part of the methodology systematic uncertainty.

\section{Systematic uncertainties}

We have evaluated several categories of systematic uncertainties using fully simulated events: uncertainties associated with jet reconstruction, jet energy measurement, $b$ tagging, the modeling of background and signal events, PDFs, and procedures and assumptions made in the analysis.
The sources of systematic uncertainties and their contributions are listed in Table~\ref{tab:syst} and added in quadrature for the total uncertainty. Details about the evaluation of the uncertainties can be found in Refs.~\cite{bib:diffxsec,Abazov:2014cca}. Additionally, we assign an uncertainty in modeling the invariant mass of the \ttbar system ($m_{\ttbar}$) based on the difference in $m_{\ttbar}$ distributions in our signal MC and the NNLO predictions~\cite{Czakon:2016ckf}.
\begin{table}[h!]
\begin{centering}
\begin{ruledtabular}\begin{tabular}{lccc}
Source & Beam & Helicity & Transverse \\
\hline
 ~~Jet reconstruction			& $\pm0.010$ & $\pm0.008$ & $\pm0.008$ \\
 ~~Jet energy measurement		& $\pm0.010$ & $\pm0.023$ & $\pm0.006$ \\
 ~~$b$ tagging				& $\pm0.009$ & $\pm0.014$ & $\pm0.005$ \\
 ~~Background modeling			& $\pm0.007$ & $\pm0.021$ & $\pm0.004$ \\
 ~~Signal modeling			& $\pm0.016$ & $\pm0.020$ & $\pm0.008$ \\ 
 ~~PDFs					& $\pm0.013$ & $\pm0.011$ & $\pm0.003$ \\
 ~~Methodology				& $\pm0.013$ & $\pm0.007$ & $\pm0.009$ \\
\hline 
~~Total systematic uncertainty 		& $\pm0.030$ & $\pm0.042$ & $\pm0.017$ \\
~~Statistical uncertainty		& $\pm0.046$ & $\pm0.044$ & $\pm0.030$ \\
~~Total uncertainty			& $\pm0.055$ & $\pm0.061$ & $\pm0.035$ \\
\end{tabular}\end{ruledtabular}
\par\end{centering}
\caption{
\label{tab:syst}
Summary of the uncertainties in the measured top quark polarization along three axes. The systematic uncertainty source indicates the difference in polarization when the measurement is repeated using alternative modeling, after applying uncertainties from the employed methods, or from assumptions made in the measurement. The uncertainties are added in quadrature to form groups of systematic sources and the total uncertainty.
}
\end{table}

\section{Results}

\begin{figure}[h!]
 \begin{centering}
 \includegraphics[width=1.0\linewidth]{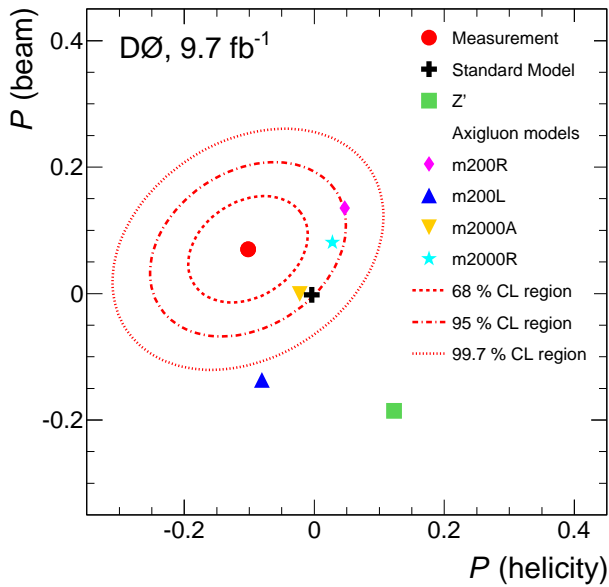} \\
 \par\end{centering}
 \caption{
 \label{fig:final}
Two-dimensional visualization of the longitudinal top quark polarizations in the $\ell$+jets channel measured along the beam and helicity axes compared with the SM and the BSM models described in the text. In this case, the m200A model is not shown as it is indistinguishable from m2000A model. 
  The correlation of the two measurement uncertainties is $27 \%$.
 }
\end{figure}

The measured polarizations for the three spin-quantization axes are shown in Table~\ref{Tab:final}. Results on the longitudinal polarizations are presented in Fig.~\ref{fig:final} and compared to SM predictions and several of the BSM models discussed previously. The measurement along the beam axis is consistent with the previous D0 result in the dilepton channel~\cite{Boris}, $P = 0.113 \pm 0.093$. 
We estimate the correlation between this result for the beam axis and that of Ref.~\cite{Boris} to be $5 \%$. The combination using the method of Refs.~\cite{Lyons:1988rp,Valassi:2003mu} yields a top quark polarization along the beam axis $P = 0.081 \pm 0.048$.
\begin{table}[h!]
\begin{center}
\begin{ruledtabular}\begin{tabular}{l c c}
Axis    & Measured polarization & SM prediction \\ \hline
Beam             & $+0.070 \pm 0.055$ & $-0.002$    \\
\,\textit{Beam - D0 comb.}             & $+0.081 \pm 0.048$ & $-0.002$    \\
Helicity             & $-0.102 \pm 0.061$ & $-0.004$    \\
Transverse             & $+0.040 \pm 0.035$ & $+0.011$    \\
\end{tabular}\end{ruledtabular}
\end{center}
\caption{
  Measured top quark polarization from the \ttbar $\ell$+jet channel along the beam, helicity, and transverse axes, and the combined polarization for beam axis with the dilepton result by the D0 Collaboration denoted as \textit{Beam - D0 comb.}. The total uncertainties are obtained by adding the statistical and systematic uncertainties in quadrature. 
}
\label{Tab:final}
\end{table}

\section{Conclusion}

In summary, we measure the top quark polarization for \ttbar production in \ppbar collisions at $\sqrt{s}=1.96$\,TeV along several spin-quantization axes. The polarizations are consistent with SM predictions. The transverse polarization is measured for the first time.  
These are the most precise measurements of top quark polarization in \ppbar collisions.

%***********************************************************

\section{Acknowledgments}

We express our appreciation to Helen Edwards for her role in designing and building the Tevatron, and her oversight of the D0 detector project in its early days.
We thank R. M. Godbole and W. Bernreuther for enlightening discussions.
We thank the staffs at Fermilab and collaborating institutions,
and acknowledge support from the
Department of Energy and National Science Foundation (USA);
Alternative Energies and Atomic Energy Commission and
National Center for Scientific Research/National Institute of Nuclear and Particle Physics  (France);
Ministry of Education and Science of the Russian Federation, 
National Research Center ``Kurchatov Institute" of the Russian Federation, and 
Russian Foundation for Basic Research  (Russia);
National Council for the Development of Science and Technology and
Carlos Chagas Filho Foundation for the Support of Research in the State of Rio de Janeiro (Brazil);
Department of Atomic Energy and Department of Science and Technology (India);
Administrative Department of Science, Technology and Innovation (Colombia);
National Council of Science and Technology (Mexico);
National Research Foundation of Korea (Korea);
Foundation for Fundamental Research on Matter (Netherlands);
Science and Technology Facilities Council and The Royal Society (United Kingdom);
Ministry of Education, Youth and Sports (Czech Republic);
Bundesministerium f\"{u}r Bildung und Forschung (Federal Ministry of Education and Research) and 
Deutsche Forschungsgemeinschaft (German Research Foundation) (Germany);
Science Foundation Ireland (Ireland);
Swedish Research Council (Sweden);
China Academy of Sciences and National Natural Science Foundation of China (China);
and
Ministry of Education and Science of Ukraine (Ukraine).


\begin{thebibliography}{99}
\bibitem{foot}
The pseudorapidity is defined as $\eta = - \ln [\tan(\theta/2)]$, where $\theta$ is the measured polar angle of an object.
\end{thebibliography}
\end{document}